Manipulation of Inverse Spin Hall Effect in Palladium by Absorption of Hydrogen Gas

S. Watt[1], M. Kostylev[1]

[1]*School of Physics and Astrophysics, University of Western Australia, Crawley, W.A. 6009, Australia*

The spintronic properties of a palladium thin film have been investigated in the presence of hydrogen gas in cobalt/palladium bilayers. Measurements of the inverse spin Hall Effect (ISHE) using cavity ferromagnetic resonance allow estimations of the spin Hall conductivity and spin diffusion length in both nitrogen and hydrogen gas atmospheres. Unwanted spin rectification effects are removed using a simple method of inverting the spin current direction with respect to the measurement setup. Absorption of hydrogen gas in the Pd layer at just 3% concentration results in a reduced ISHE voltage amplitude and bilayer resistance. Fitting the ISHE voltage against the Pd layer thickness demonstrates that the spin diffusion length decreases by 23% in the presence of a hydrogen gas. On the other hand, the results indicate that there is no significant change in the spin Hall conductivity of Pd due to hydrogen absorption.

Spintronic effects originating from spin-orbit coupling (SOC) in metallic thin films have attracted a lot attention recently. Palladium is known to exhibit a significant SOC and has been used in many experiments to measure the inverse spin Hall effect (ISHE).[1–6] In addition, cobalt/palladium (Co/Pd) bilayers are known to possess a strong interfacial perpendicular anisotropy (PMA) which modifies the ferromagnetic resonance (FMR) properties of the film.[7] Unique to Pd is its ability to reversibly absorb hydrogen gas ($H_2$) at room temperature and atmospheric pressure that is widely used in sensing applications. Chemical transformation of the pure Pd metal into Pd hydride takes place on absorption which alters the PMA[8–10] resulting in a shift in the resonance condition.[11] This shift has been shown to depend on $H_2$ concentration in the film environment and sensitive in the whole concentration range.[12,13]

In our recent paper[14], we found that the ISHE voltage peak measured in the conditions of FMR in the Co layer exhibits the same resonance field shift as the simultaneously measured FMR peak in response to exposure of a Pd/Co film to $H_2$. The shift is accompanied by a linewidth decrease suggesting a modification of the strength of ISHE in Pd in the presence of hydrogen. This study aims to quantify the effect of incorporation of hydrogen atoms into Pd lattice on the strength of ISHE.

This study delivers more insight into the physics underlying the ISHE effect enabled by the ability of Pd to reversibly change its physical properties by absorbing hydrogen. Accordingly, taking measurements of the ISHE voltage in Pd for different $H_2$ concentrations is equivalent to characterising a set of separate samples, all with different material properties yet having the same fabrication history. Furthermore, due to reversibility of the gas absorption by Pd, the experiment can be repeated multiple times without need to fabricate new samples.

This investigation is also important for future applications. $H_2$ sensors based on magnetisation dynamics in thin films show great promise for safe, efficient and sensitive $H_2$ detectors. With increased attention being given to hydrogen fuel usage and storage as an alternative energy source, there is an unsatisfied demand for more versatile sensors.[15] A recent sensor prototype which has demonstrated stable and efficient detection of $H_2$ at a wide range of concentration levels utilises the FMR peak shift of Co/Pd films in the presence of hydrogen.[11,16] Accordingly, the recently observed modification of the ISHE response of the bilayers in the presence of $H_2$ opens up new avenues for designing $H_2$ detectors.

Here we investigate the effect of $H_2$ on the spintronic properties of Pd, exploring the effect of hydrogen on the spin Hall conductivity ($\sigma_{Pd}^{SH}$) and spin diffusion length ($\lambda_{sd}$). While quantification of these quantities has been undertaken already by many studies,[1–6,17–20] the aim of this experiment is to see how they are modified in the presence of $H_2$.

Induced ISHE voltages are observed across ferromagnetic/nonmagnetic bilayers via the process of spin pumping.[1–6,19–34] Dynamically driven magnetisation precession produces a spin current which is converted to a DC charge current by the ISHE. The amplitude of the pumped spin current scales as the precession angle and reaches a maximum during FMR resulting in a Lorentzian voltage lineshape as a function of applied magnetic field $\vec{H}$.

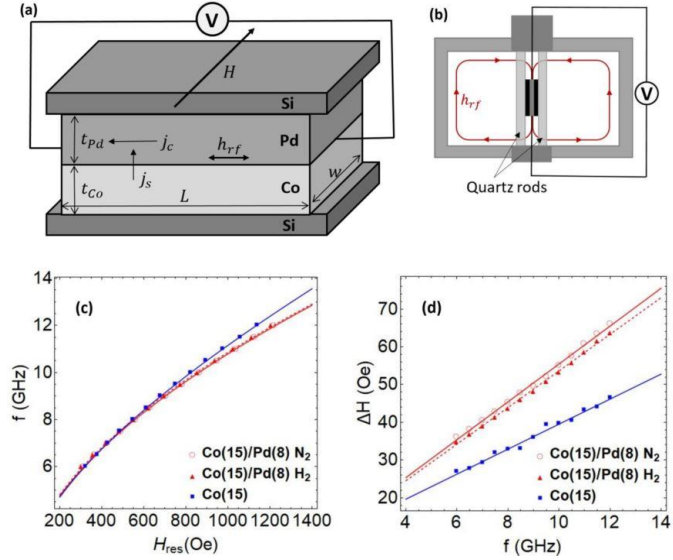

FIG. 1. (a) Sample schematic. (b) Side profile of the microwave cavity with $\vec{H}$ into the page. The quartz holder and sample are symmetric with respect to the cavity. (c) $H_r$ versus frequency for reference and Co(15 nm)Pd(8 nm) samples. Curves are fit with Eq. (2) yielding $\gamma = 2.82$ MHz/Oe and $4\pi M_{eff} = 15390$ Oe for the reference. (d) Linewidth against frequency for reference and Co(15nm)Pd(8nm) samples. Fits are to $\Delta H = \Delta H_0 + \alpha \frac{2\pi f}{\gamma}$.

Difficulty in measuring ISHE arises due to spurious spin rectification effects (SRE) which can couple microwave currents in the ferromagnetic layer to the dynamic magnetisation to produce DC voltages. Anisotropic magnetoresistance (AMR) and anomalous Hall effect (AHE) are commonly observed during FMR, with resulting DC voltages proportional to $(\vec{J}_{rf} \cdot \vec{M})\vec{M}$ and $(\vec{J}_{rf} \times \vec{M})$ respectively.[35] Here $\vec{J}_{rf}$ is the microwave current induced in the ferromagnetic layer due to the microwave field in which the film is placed and $\vec{M}$ is the total magnetisation vector for the ferromagnetic material.

The SRE voltages $V_{SRE}$ can have similar lineshape and amplitude to the ISHE voltage $V_{ISHE}$ during FMR,[35,36] and thus careful separation is required to extract $V_{ISHE}$. Both the electric and magnetic components of the driving microwave field can induce microwave currents in the sample.

To minimise the effect of the SRE on the measured DC voltage, we employ a microwave cavity to drive FMR. We place the sample in the cavity centre where the microwave electric field is vanishing. To eliminate any residual $V_{SRE}$, we use symmetry arguments similar to Zhang et al.[34] to extract the ISHE voltage by a film-flip subtraction technique. ISHE converts a spin current ($j_s$) flowing in Pd to a charge current according to $\vec{J_c} = \theta_{SHE} j_s (\vec{n} \times \vec{\sigma})$, where $\vec{\sigma}$ is the polarisation vector of the electron spins pumped into Pd, $\vec{n}$ is the direction of the pumped spin current (normal to the Co/Pd interface) and $\theta_{SHE}$ is the spin Hall angle. By inverting the spin current direction with respect to the spin polarisation vector, the measured charge current due to the ISHE is inverted. This can be achieved either by reversing the magnetic field (inverting $\vec{\sigma}$) or flipping the film by 180° (inverting $\vec{n}$). We employ the latter case where the film is flipped upside down while keeping $\vec{H}$ fixed. In this scenario, $\vec{M}$ will remain unchanged with respect to the microwave currents in the sample while $V_{SRE}$ will remain constant.

Thus by flipping the sample and keeping the measurement setup fixed, only $V_{ISHE}$ is inverted. This technique requires the assumption that the microwave currents induced in the conductive film are the same regardless of the film orientation. To help ensure this, the films are symmetrised so that flipping the samples does not modify the microwave fields which drive FMR. Based on the above argument, $V_{SRE}$ and $V_{ISHE}$ are easily separated as

$$V_{ISHE/SRE} = \frac{1}{2}(V_{up} \pm V_{down}), \quad (1)$$

Films of Co and Pd are sequentially sputter deposited onto silicon substrates. Layer thicknesses are set to 15 nm for Co while the Pd thickness is varied from 3-35 nm. A single-layer Co film is deposited for reference. After deposition, the samples are cut to dimensions of width, $w = 4\,mm$ and length, $l = 10\,mm$. Electrical contacts are made to the samples' short edges using a silver conductive epoxy. The samples are symmetrised by placing pieces of bare substrate of identical dimensions on top of the bilayer, effectively sandwiching the bilayer in symmetric stack (see Fig 1 (a)). Each sample is also measured using in-plane (IP) stripline FMR.

For cavity FMR measurements, the samples are loaded into a TE102 cavity on a quartz holder. The sample is placed at the cavity centre where the microwave electric field is minimum and the microwave magnetic field is maximum and parallel to the electrical contacts. The cavity has a resonance frequency of 9.5 GHz and a loaded Q factor of 2720. The quartz holder is symmetric with respect to the sample and to the cavity (see Fig 1 (b)) such that a rotation of 180 degrees does not modify the resonant mode in the cavity. The bilayer is magnetised IP and perpendicular to the electrical contacts. Microwave power is incident to the cavity at 158 mW through a rectangular waveguide and the coupling is tuned using an iris screw so the reflected power is zero. The microwave magnetic field at the cavity centre is estimated as $h_{rf} = 0.58$ Oe.

The voltage across the sample length is measured using a Keysight 34420A nanovoltmeter as a function of $\vec{H}$. Field sweeps are performed for the initial sample orientation ($V_{up}$) then the sample holder is rotated by 180 degrees such that the spin current is reversed with respect to the $\vec{H}$ and the cavity. The field sweep is performed again ($V_{down}$). Due to the symmetry of the sample stack and sample holder, the cavity mode distribution is identical for both sweeps and the induced microwave currents in the sample due to the dynamic driving field remain unchanged with respect to $\vec{H}$ and so $V_{SRE}$ remains unchanged.

For hydrogenation measurements, the samples are first measured in nitrogen gas at atmospheric pressure followed by a 3%/97% hydrogen/nitrogen gas mixture, allowing 30 minutes to hydrogenate fully. Previous studies have shown that concentrations down to 0.2% can be detected using the FMR based H$_2$ sensing method in similar bilayer samples.[12,13]. The number density of hydrogen atoms to Pd atoms should be slightly less than 30% in the 3% H$_2$ partial pressure state.[37]

Examples of the IP stripline measurements are shown in Fig 1 (c-d) for the reference Co layer and the Co(15 nm)Pd(8 nm) bilayer. The resonance field $H_r$, measured at different frequencies can be used to extract the gyromagnetic ratio, $\gamma$, by fitting to the IP Kittel equation:

$$\omega = \gamma\sqrt{H_r(H_r + 4\pi M_{eff})}. \quad (2)$$

Here $4\pi M_{eff}$ is the effective magnetisation. Pd causes an interface PMA when layered with a ferromagnetic material,[7] resulting in a reduction of $4\pi M_{eff}$. This is evident in Fig 1 (c) where for a given frequency, $H_r$ are shifted to higher values. Absorption of H$_2$ results in a further modification to the interface PMA in Co.[8–10]

Fig 1 (d) shows the FMR half-linewidth, $\Delta H$ as a function of frequency. Fitting to $\Delta H = \Delta H_0 + \alpha \frac{2\pi f}{\gamma}$ allows extraction of the damping parameter $\alpha$ and the inhomogeneous linewidth broadening, $\Delta H_0$. A stark difference between the single layer Co and the bilayer case is observed as a result of the enhanced damping due to spin pumping.[38,39] Also evident is the decrease of about 2% in $\alpha$ in the bilayer samples upon hydrogenation which is consistent with previous work.[40]

Fig 2 (a) shows a typical set of cavity FMR measurements in nitrogen atmosphere for the Co(15 nm)Pd(8 nm) sample. While the raw traces for $V_{up}$ and $V_{down}$ are almost Lorentzian in shape, they do exhibit some asymmetry due to the presence of $V_{SRE}$. Each raw trace is fitted with a combination of symmetric Lorentzian and antisymmetric dispersive lineshapes.[31] After performing the subtraction between the two measurements, one finds that the voltage has an almost perfect Lorentzian lineshape as expected from a spin pumping origin.

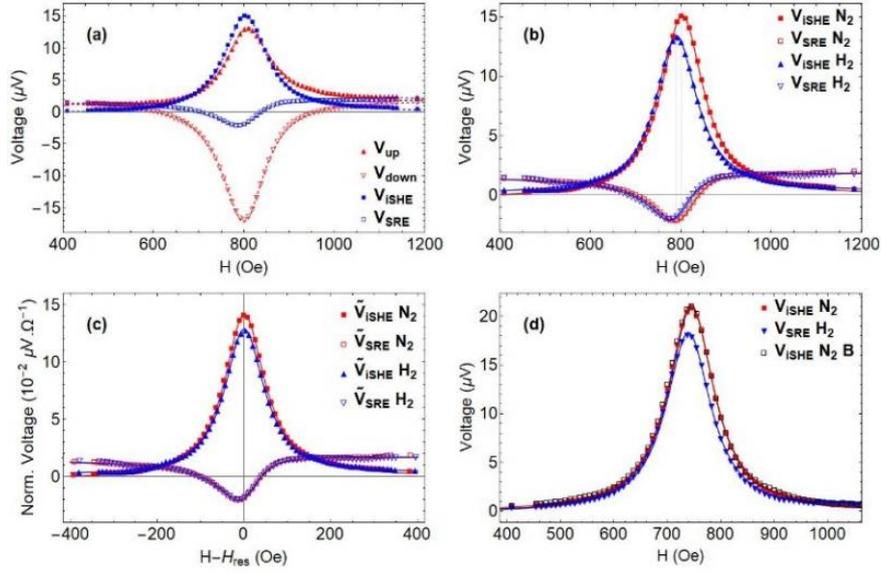

FIG. 2. (a) Raw voltage traces (red) before and after flipping the sample and the extracted $V_{ISHE}$ and $V_{SRE}$ (blue). (b) ISHE and SRE voltages for the Co(15 nm)Pd(8 nm) sample in nitrogen and hydrogen atmospheres. (c) $V_{ISHE}$ and $V_{SRE}$ centred and normalised by the bilayer resistance. (d) $V_{ISHE}$ before, during and after hydrogenation.

Comparison of $V_{ISHE}$ to $V_{SRE}$ shows that the former is the dominant contribution in these films. Changes to the cavity loading due to variations in sample positioning and wiring configurations would result in differences in the incident power to the sample. These small variations are difficult to avoid and may result in small changes in both voltages when rotating the sample holder. Confidence in the results comes from the fact that the processed ISHE traces are almost perfectly symmetric with a consistently low noise level.

Fig 2 (b) shows the effect of exposing the $H_2$ mixture on the Co(15 nm)Pd(8 nm) sample which has multiple effects on the sample, the most significant of which is the decrease in the $V_{ISHE}$ amplitude (the main focus of this study). Hydrogenation of the bilayers also results in a slight decrease in total electric resistance. The electric resistivity of single-layer Pd grown on Si substrates is known to increase upon hydrogenation due to an increase in the scattering of the electrical carriers.[41] The situation appears to be different for the bilayer case, which we speculate is due to a lessening of the interfacial compressive stress caused by the smaller unit cell of Co with respect to Pd. Hydrogen absorption in Pd is known to cause expansion of the lattice which may result in a decreased resistivity, overcompensating the increased scattering off the hydrogen atoms. It should be noted that the decrease in resistance is not sufficient to result in the $V_{ISHE}$ amplitude change alone, which scales as the total bilayer resistance $R_T$. Fig 2 (c) shows $V_{ISHE}$ centred and normalised by $R_T$, showing that the amplitude decrease of $V_{ISHE}$ cannot be explained by the bilayer resistance decrease alone. One may also note in Fig 2 (b) a small change in the amplitude of $V_{ISHE}$. This is typical for all samples and *can* be explained by the change in the resistance of the Pd layer, which changes the total bilayer resistance. As shown in Fig 2 (c), the normalised $V_{SRE}$ overlay almost perfectly. This is not surprising given that $V_{SRE}$ originates in the Co layer alone.

By assuming a shunt relationship between the resistances of each layer, the conductivity can be calculated as shown in Fig 3 (a). We calculate $\sigma_{Pd} = (1.47 \pm 0.10) \times 10^6 \, (\Omega m)^{-1}$ which increases by 3.7% upon hydrogenation. This value can be used to calculate the spin Hall conductivity by $\sigma_{Pd}^{SH} = \sigma_{Pd}\theta_{SHE}$. The Co layer conductivity is assumed constant.

After hydrogenation, the cavity is flushed with $N_2$ and the Pd releases its stored hydrogen. Fig 2 (d) shows $V_{ISHE}$ before, during and after hydrogenation. Both $R_T$ and $V_{ISHE}$ return to the pre-hydrogenation values indicating the observed effects are related to the presence of hydrogen in the Pd lattice and not irreversible structural changes occurring in Pd.

In order to determine $\theta_{SHE}$ and $\lambda_{sd}$, the thickness dependent $V_{ISHE}$ is fit to (following Mosendz *et al.*[5])

$$V_{ISHE} = few \, R_T P \sin^2(\Phi) \, g_{eff}^{\uparrow\downarrow} \, \theta_{SHE}\lambda_{sd} \tanh\left(\frac{t_{Pd}}{2\lambda_{sd}}\right). \quad (3)$$

Here $e$ is the electron charge and $\Phi$ is the cone angle of the magnetisation precession determined by $\Phi = \frac{h_{rf}}{\Delta H}$. The factor P accounts for ellipticity of the precession cone which modifies the pumped spin current. $V_{ISHE}$ is proportional to the real part of the effective spin mixing conductance $g_{eff}^{\uparrow\downarrow}$ which is determined using the effective damping parameters extracted from the IP FMR $\Delta H$ measurements. Finally, we normalise $V_{ISHE}$ by all known values

$$V_{ISHE}^{norm} = \frac{V_{ISHE}\,\Delta H^2}{fewR_T P\, h_{rf}^2 g_{eff}^{\uparrow\downarrow}} = \theta_{SHE} \lambda_{sd} \tanh\left(\frac{t_{Pd}}{2\lambda_{sd}}\right). \quad (4)$$

Plots of $V_{ISHE}^{norm}$ against the Pd thickness are shown in Fig 3 (b). Despite the significant uncertainty in the data, there is a clear reduction in $V_{ISHE}^{norm}$, indicative of a change in either $\theta_{SHE}$ or $\lambda_{sd}$. From the fits to Eq. (4), the extracted $\theta_{SHE}$ values of $(2.17 \pm 0.19)\%$ and $(2.21 \pm 0.25)\%$ in nitrogen and hydrogen atmospheres respectively. The small increase in $\theta_{SHE}$ is less than the experimental uncertainty in the values however, and so the hydrogen cannot be said to have had any significant effect. Due to a significant spread in the estimated value of $\theta_{SHE}$ of Pd in the literature ($\theta_{SHE} = 0.48\%$ from Tao *et al.*[4] to $\theta_{SHE} = 10\%$ from Kumar *et al.*[19]), the value estimated in this study is certainly reasonable. Using our estimated $\theta_{SHE}$, we calculate $\sigma_{Pd}^{SH} = (319 \pm 37)(\Omega cm)^{-1}$ and $\sigma_{Pd}^{SH} = (337 \pm 46)(\Omega cm)^{-1}$ in nitrogen and hydrogen respectively, in agreement with theoretical $(350\ (\Omega cm)^{-1}$ from Guo[17]), and experimental $(250\ (\Omega cm)^{-1}$ from Tao *et al.*[4]) literature values for $\sigma_{Pd}^{SH}$ in air.

Fitted values for $\lambda_{sd}$ are $(5.79 \pm 0.84)nm$ and $(4.47 \pm 0.73)nm$ for nitrogen and hydrogen atmospheres respectively. Literature values for $\lambda_{sd}$ are in good agreement.[4,19,20] The 22.8% decrease in spin diffusion length, above the uncertainty, in the presence of hydrogen is likely due to an increase in spin scattering off the hydrogen atoms causing the electrons to lose polarisation faster. To determine the dependence of $\lambda_{sd}$ on differing $H_2$ concentrations, further experiments are needed since Pd can readily absorb $H_2$ and changes in the FMR properties can be observed up to 100% concentration.[12,13]

To summarise, the spintronic properties of a Pd thin film have been investigated in the presence of $H_2$ by use of ISHE. Estimations of the spin Hall conductivity and spin diffusion length based on fitted data give reasonable values in agreement with literature. Absorption of hydrogen did not result in any significant change in the spin Hall conductivity. The spin diffusion length was shown to decrease by approximately 23% in the presence of $H_2$. These results may prove useful for the development of a spintronic based $H_2$ sensor while also being of use in the study of spin Hall phenomena in condensed matter physics.

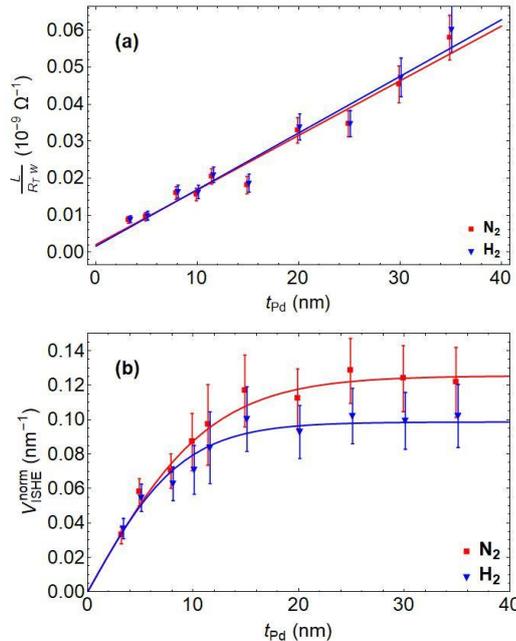

FIG. 3. (a) Bilayer conductivity with linear fits using $\frac{L}{R_T w} = (\sigma_{Pd} t_{Pd} + \sigma_{Co} t_{Co})$. (b) $V_{ISHE}^{norm}$ for both atmospheres with fits to $V_{ISHE}^{norm} = \theta_{SHE} \lambda_{sd} \tanh\left(\frac{t_{Pd}}{2\lambda_{sd}}\right)$.

Acknowledgements

Research Collaboration Award and Vice Chancellor's Senior Research Award from the University of Western Australia are acknowledged. The authors would also like to acknowledge the facilities, and the scientific and technical assistance of the Australian National Fabrication Facility, Centre for Microscopy, Characterisation and Analysis, The University of Western Australia, a facility funded by the University, State, and Commonwealth Governments. The work of S. Watt was supported by the Australian Government Research Training Program.